\documentclass[aip,rsi,reprint,graphicx]{revtex4-1}

\usepackage{color}

\newcommand{\grad}{\nabla}

\newcommand{\KAISTNQe}{Department of Nuclear and Quantum Engineering, KAIST, Daejeon 34141, Republic of Korea}

\newcommand{\MaxPlanckGreifswald}{Max-Planck-Institut f$\ddot{u}$r Plasmaphysik, Teilinstitut Greifswald, D-17491 Greifswald, Germany}

\newcommand{\NFRI}{National Fusion Research Institute, Daejeon 34133, Republic of Korea}

\newcommand{\smjoung}{\author{Semin~Joung}\email[]{smjoung@kaist.ac.kr}\affiliation{\KAISTNQe}}
\newcommand{\jwkim}{\author{Jaewook~Kim}\affiliation{\KAISTNQe}}
\newcommand{\krpark}{\author{Kyeo-reh~Park}\affiliation{\KAISTNQe}}

\newcommand{\ycghim}{\author{Y.-c.~Ghim}\email[]{ycghim@kaist.ac.kr}\affiliation{\KAISTNQe}}
\newcommand{\shkwak}{\author{Sehyun~Kwak}\affiliation{\KAISTNQe}\affiliation{\MaxPlanckGreifswald}}

\newcommand{\shhahn}{\author{S.H.~Hahn}\affiliation{\NFRI}}
\newcommand{\hshan}{\author{H.S.~Han}\affiliation{\NFRI}}
\newcommand{\hskim}{\author{H.S.~Kim}\affiliation{\NFRI}}
\newcommand{\jgbak}{\author{J.G.~Bak}\affiliation{\NFRI}}
\newcommand{\sglee}{\author{S.G.~Lee}\affiliation{\NFRI}}

\expandafter\let\csname equation*\endcsname\relax
\expandafter\let\csname endequation*\endcsname\relax
\usepackage[toc,page]{appendix}
\usepackage{graphicx}
\usepackage{caption}
\usepackage{amstext}
\usepackage{amssymb}
\usepackage{amsmath}
\usepackage[utf8]{inputenc}
\usepackage{booktabs}
\usepackage{array}
\usepackage{multirow}
\usepackage{kotex}
\usepackage{amsbsy}
\usepackage{subfig}
\usepackage{hyperref}
\usepackage{accents}
\newcommand\munderbar[1]{%
  \underaccent{\bar}{#1}}
\hypersetup{
    colorlinks=true,
    linkcolor=blue,
    citecolor=blue,
}

\newcommand{\boldlambda}{\boldsymbol{\lambda}}
\newcommand{\boldB}{\boldsymbol{B}}
\newcommand{\boldX}{\boldsymbol{X}}
\newcommand{\boldx}{\boldsymbol{x}}
\newcommand{\boldL}{\boldsymbol{L}}
\newcommand{\matK}{\boldsymbol{\munderbar{\munderbar{K}}}}
\newcommand{\matLambda}{\boldsymbol{\munderbar{\munderbar{\Lambda}}}}

\makeatletter
\newcommand*{\rom}[1]{\expandafter\@slowromancap\romannumeral #1@}
\makeatother

\begin{document}

\title{Bayesian with Gaussian process based missing input imputation scheme for reconstructing magnetic equilibria in real time}
\thanks{Contributed paper published as part of the Proceedings of the 22nd Topical Conference 
on High-Temperature Plasma Diagnostics, San Diego, California, April, 2018.\\}

\smjoung
\jwkim
\shkwak
\krpark
\shhahn
\hshan
\hskim
\jgbak
\sglee
\ycghim

\date{\today}
\begin{abstract}
A Bayesian with GP(Gaussian Process)-based numerical method to impute a few missing magnetic signals caused by impaired magnetic probes during tokamak operations is developed such that the real-time reconstruction of magnetic equilibria, whose performance strongly depends on the measured magnetic signals and their intactness, are affected minimally. Likelihood of the Bayesian model constructed with the Maxwell's equations, specifically Gauss's law of magnetism and Amp\`{e}re's law, results in infinite number of solutions if two or more magnetic signals are missing. This undesirable characteristic of the Bayesian model is remediated by coupling the model with the Gaussian process. Our proposed numerical method infers the missing magnetic signals correctly in less than $1$\:msec suitable for real-time reconstruction of magnetic equilibria during tokamak operations. The method can also be used for a neural network that reconstructs magnetic equilibria trained with a complete set of magnetic signals. Without our proposed imputation method, such a neural network would become useless if missing signals are not tolerable by the network.
\end{abstract}

\maketitle

%
\vspace{2pc}
%
%
%
%

\section{Introduction}

Magnetic pick-up coils installed on magnetic confinement devices such as tokamaks and stellarators in addition to Rogowski and flux loop coils provide magnetic information such that high temperature fusion-grade plasmas can be controlled in real time and that magnetic equilibria can be reconstructed for data analyses. Neural networks, also, have been developed to provide the positions of X-point and plasma boundary in real time\cite{Lister:1991gx, Coccorese:1994jt} where input signals to the networks are magnetic signals. Therefore, integrity and intactness of the magnetic signals are of paramount importance; nevertheless magnetic probes are susceptible to impairments during plasma operations, resulting in missing or specious magnetic signals whose consequences may include faulty plasma operations and incorrect data analyses. For the case of neural networks trained with full sets of magnetic signals, even a single missing signal may cause the networks not to work properly.

We present how one can numerically infer, thus impute, missing magnetic signals in real time based on a Bayes' model\cite{Sivia:2006} coupled with the Gaussian Process\cite{Rasmussen:2006} (GP). Likelihood is constructed based on the Maxwell's equations, specifically Gauss's law of magnetism and Amp\`{e}re's law, consistent with the measured data. A couple of algorithms to detect faulty magnetic sensors have been developed,\cite{Neto:2014cm, Nouailletas:2012ho} and an inference method for just one faulty signal has also been proposed.\cite{Nouailletas:2012ho} Our proposed method in this work is examined with up to nine missing magnetic probe signals installed on KSTAR,\cite{Kwon:2011dq} and we find that the method infers the correct values in less than $1$ msec on a typical personal computer suitable for the real-time plasma operations, real-time EFIT reconstruction\cite{Lao:1985hn, Ferron:2002fw} and neural networks trained with complete sets of magnetic signals. We note that detecting faulty or missing magnetic signals can be done as suggested elsewhere.\cite{Neto:2014cm, Nouailletas:2012ho}

Detailed descriptions on how we generate the likelihood and estimate the maximum a posterior of the Bayes' model and how well the model infers the missing values as well as its limitation are provided in section \ref{subsec:bayes}. The limitation on the Bayes' model motivats us to use the GP discussed in section \ref{subsec:GP} which also has a certain drawback. In section \ref{subsec:bayes_GP}, we present surpassing performance, i.e., resolving the defects of the Bayes' model and the GP while retaining their advantages, achieved by coupling the Bayes' model with the GP. To examine our proposed method we assume that the intact magnetic signals are missing and compare the measured signals with the inferred values. Our conclusion is presented in section \ref{sec:conclusion}.

\section{Imputation Scheme}
\label{sec:imputation}

\subsection{Based on Bayes' model}
\label{subsec:bayes}

\begin{figure}[t]
    \centering
    \includegraphics[width=0.99\linewidth]{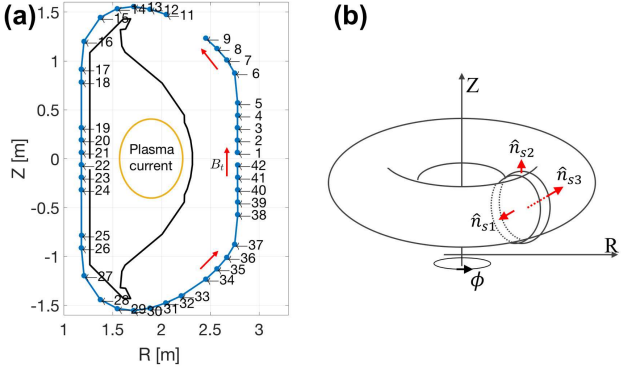}
    \caption{Schematics of (a) the Amperian loop (blue line connecting blue dots) for $\grad\times\vec B = \mu_0\vec J$ and (b) the pancake-shaped Gaussian surface with three surfaces $s_1$, $s_2$ and $s_3$ for $\grad\cdot\vec B = 0$. Blue dots with the numbers in (a) indicate the magnetic probes at a certain toroidal location of KSTAR. $\hat n$ is a unit normal vector, and we generate the Gaussian surface such that $\hat n_{s_1}\cdot\hat n_{s_3}=-1$, while $n_{s_2}$ is parallel to $B_n$.}
    \label{fig:MS-modelling}
\end{figure}

Magnetic probes, depicted in Fig. \ref{fig:MS-modelling}(a) as the blue dots with the probe numbers, installed on KSTAR\cite{Lee:2008cl} at a certain toroidal location measure tangential ($B_t$) and normal ($B_n$) components of the magnetic fields with respect to the wall. Missing tangential components are inferred with Amp\`{e}re's law, i.e., $\grad\times\vec B=\mu_0\vec J$ neglecting $\partial\vec E/\partial t$ term based on a usual magnetohydrodynamic assumption,\cite{Freidberg:2014dt} and missing normal components with Gauss's law of magnetism, i.e., $\grad\cdot\vec B=0$.

With the Amperian loop, the blue line connecting the blue dots shown in Fig. \ref{fig:MS-modelling}(a), the tangential components of the magnetic signals $B_t$ must approximately satisfy
\begin{eqnarray}
\label{eq:ampere}
\mu_0 I_p &=& \oint_L \vec B \cdot d\vec{l} \approx \oint_L \left( B_t^\text{MP} - B_t^\text{PF} \right) \: dl \nonumber \\
&\approx& \left[ \sum_{m=1}^{N_m} \Delta l_{m}^* \left( B_{t, m}^{*\text{MP}} - B_{t, m}^{*\text{PF}} \right) + \right. \nonumber \\
&& \qquad\qquad\qquad\qquad \left. \sum_{i=1}^{N_i} \Delta l_i \left( B_{t, i}^\text{MP} - B_{t, i}^\text{PF} \right) \right] \nonumber \\
&=&\boldlambda^{*T}\left(\boldB_t^{*\text{MP}} - \boldB_t^{*\text{PF}} \right) + \nonumber \\
&& \qquad\qquad\qquad\qquad \boldlambda^T \left(\boldB_t^{\text{MP}} - \boldB_t^{\text{PF}} \right), 
\end{eqnarray}
where $I_p$ is the total plasma current. $B_t^\text{MP}$ and $B_t^\text{PF}$ are the tangential components of the magnetic fields measured by the magnetic probes (MP) and induced by the poloidal field (PF) coils, respectively. Note that KSTAR has 14 PF coils, and as the magnetic probes also sense the magnetic fields induced by the PF coils we need to remove such effects\cite{Tsaun:2007il} as in $B_t^\text{MP} - B_t^\text{PF}$. $m$ and $i$ are the indices for the missing and the intact magnetic signals; whereas $N_m$ and $N_i$ are the total numbers of the missing and the intact signals, respectively. $\Delta l$ denotes the segment distance between the magnetic probes, and it is different for different probes as can be seen in Fig. \ref{fig:MS-modelling}(a). Superscripted asterisk means the missing magnetic signal. The last line in Eq. (\ref{eq:ampere}) is just a reformulation of the second line using the vector notations, i.e., $\boldlambda^{(*)}=\{\Delta l^{(*)}_{i(m)}\}$ and $\boldB_t^{(*)}=\{B_{t,i(m)}^{(*)}\}$. Moret \textit{et al.}\cite{Moret:1998fh} has used Eq. (\ref{eq:ampere}) to obtain plasma currents in TCV tokamak; whereas we apply the same idea to obtain the missing magnetic signals based on the plasma currents measured by Rogowski coils.

For the normal components of the magnetic signals, we utilize the pancake-shaped Gaussian surface as depicted in Fig. \ref{fig:MS-modelling}(b) consisting of three surfaces $s_1$, $s_2$ and $s_3$. We assume (or force) that the Gaussian surface is flat enough, so that the magnetic fluxes through the surfaces of $\hat n_{s_1}$ and $\hat n_{s_3}$ cancel each other as $\hat n_{s_1}\cdot\hat n_{s_3}=-1$, where $\hat n$ is a unit normal vector. Then, $\grad\cdot\vec B=0$ can be written as
\begin{equation}
\label{eq:gauss_1}
0=\oint_{s_1+s_2+s_3} \!\!\!\!\!\!\!\!\!\!\!\!\!\!\!\! \vec B \cdot d\vec S \approx \int_{s_2} B_n\:dA\approx\Delta w \int_L B_n\: dl,
\end{equation}
where $B_n$ is the normal component of the magnetic field and $dA$ ($= \Delta w\:dl$) the differential area normal to the surface $s_2$ (parallel to $B_n$) with $\Delta w$ being the thickness of the Gaussian surface. $dl$ is the differential length encompassing the minor radius (or the poloidal cross-section) and essentially same as the blue line in Fig. \ref{fig:MS-modelling}(a). Since $\Delta w\ne 0$, we then have, again with the vector notations, 
\begin{eqnarray}
\label{eq:gauss}
0&=&\int_L B_n\: dl=\boldlambda^{*T}\boldB_n^{*}  + \boldlambda^T \boldB_n.
\end{eqnarray}
Here, we do not need to separate out the PF coil effects since $\grad\cdot\vec B=0$ is true whatever the sources are.

Our likelihood assuming that the noise in magnetic signals are Gaussian, then, becomes
\begin{eqnarray}
\label{eq:likebi}
p( \boldB_\oplus | \boldB_\oplus^*, \Omega_\oplus) &=& \frac{1}{\sqrt{2 \pi} \sigma} \times \nonumber \\
&&\!\!\!\!\!\!\!\!\!\!\!\!\!\!\!\!\!\!\!\!\!\!\!\!\!\! \exp \left[-\frac{\left(\boldlambda^{*T} \boldB_\oplus^* - \left(\Omega_\oplus - \boldlambda^{T} \boldB_\oplus \right) \right)^2}{2 \sigma^2} \right],
\end{eqnarray}
where $\boldB_\oplus^{(*)}$ is either $\boldB_t^{(*)\text{MP}}-\boldB_t^{(*)\text{PF}}$ or $\boldB_n^{(*)}$ depending on whether we are interested in the tangential or normal component, respectively. Likewise, the value of $\Omega_\oplus$ is $\mu_0 I_p$ for the tangential component or simply $0$ for the normal component. $\sigma^2$ is the noise variance which is estimated based on the measured magnetic signals. 

\begin{figure}[t]
    \centering
    \includegraphics[width=0.95\linewidth]{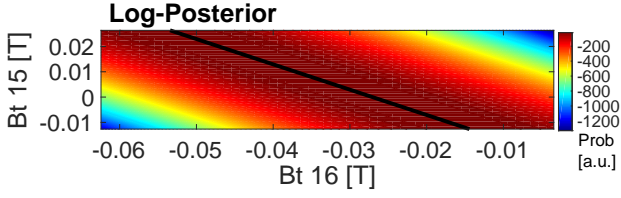}
    \caption{Log-posterior of the missing magnetic signals inferred by the Bayes' model with the Maxwell's equations, i.e., Eqs. (\ref{eq:likebi}) and (\ref{eq:msbayeinf}), when two tangential components ($B_t$ from MPs \#15 and \#16) of the magnetic signals are missing. Thick black line indicates where the posterior is maximum indicating that infinite number of solutions are possible. Data are inferred for KSTAR shot \#9010 at $0.1$\:sec.}
    \label{fig:Inference-limit}
\end{figure}

Finally, we obtain posterior as
\begin{equation} \label{eq:msbayeinf}
p(\boldB^*_\oplus | \boldB_\oplus, \Omega_\oplus) \propto p(\boldB_\oplus | \boldB_\oplus^* , \Omega_\oplus)\:p(\boldB^*_\oplus | \Omega_\oplus),
\end{equation}
providing us inferred values of the missing magnetic signals ($\boldB_\oplus^*$) consistent with the measured signals ($\boldB_\oplus$ and $\sigma$) and the Maxwell's equations ($\Omega_\oplus$). With a uniform prior $p(\boldB^*_\oplus|\Omega_\oplus)$, it is obvious that we obtain infinite number of solutions from \textit{maximum a posterior} (MAP) method if we have more than one unknowns of the same component. In simpler words, we have only one equation for the tangential (Amp\`ere's Law) or the normal (Gauss's law of magnetism) component; thus, more than one unknowns of the same component result in infinite number of solutions. Fig. \ref{fig:Inference-limit} showing an estimated log-posterior distribution where we have removed two $B_t$ measurements, i.e., probe numbers \#15 and \#16, confirms this effect clearly as depicted by the thick black line corresponding to the MAPs. This is the limitation of the imputation scheme solely based on the Bayes' model consistent with the Maxwell's equations.

\subsection{Based on Gaussian Process}
\label{subsec:GP}

Motivated by the limitation of the Baye's model with the Maxwell's equations, we introduce Gaussian Process\cite{Rasmussen:2006} (GP) in our imputation scheme. We express the probability distribution of $\boldB^*$ ($N_m \times 1$ column vector) given the measured data $\boldB$ ($N_i \times 1$ column vector) without any analytic expression of the data \textit{a priori} as described elsewhere\cite{Rasmussen:2006, Mises:1964}
\begin{equation}
\label{eq:gp}
p\left(\boldB^* | \boldB \right) = \mathcal{N}\left( \matK^*\matK^{-1}\boldB,\; \matK^{**}-\matK^*\matK^{-1}\matK^{*T}\right),
\end{equation}
with
\begin{eqnarray}
\label{eq:gpcond}
&& \quad \matK \equiv \matK\left(\boldX, \boldX \right), \quad (N_i\times N_i\text{ matrix}) \nonumber \\
&&\quad \matK^* \equiv \matK\left(\boldX^*, \boldX \right), \quad (N_m\times N_i\text{ matrix}) \nonumber \\
 &&\quad\matK^{**} \equiv \matK\left(\boldX^*, \boldX^* \right), \quad (N_m\times N_m\text{ matrix}), \nonumber
\end{eqnarray}
where $\mathcal{N}(\:,\:)$ is the usual notation for a normal distribution. Recall that $N_i (N_m)$ is the total number of intact (missing) magnetic signals. Here, $\boldX^{(*)}$ is the $2\times N_i (N_m)$ matrix containing the physical positions of all the intact (missing) magnetic probes in two dimensional space, i.e., physical $R$ and $Z$ positions at a fixed toroidal location. 

The $i^{\text{th}}$ and $j^{\text{th}}$ component of a covariance matrix $\matK^{(*\text{ or }**) }$ is defined as
\begin{eqnarray}
&& K_{ij}^{(*\text{ or }**)}\left(\boldx_i^{(*)}, \boldx_j^{(*)}\right) = \nonumber \\
&& \sigma_f^2 \exp\left[-\frac{1}{2} \left(\boldx_i^{(*)}-\boldx_j^{(*)}\right)^T  \begin{bmatrix} \ell_{R}^2 & 0 \\0& \ell_{Z}^2 \end{bmatrix}^{-1} \left(\boldx_i^{(*)}-\boldx_j^{(*)}\right)  \right] \nonumber \\
&& +\delta_{ij}\sigma_n^2,
\end{eqnarray}
where $\boldx_i^{(*)}$ is the $i^{\text{th}}$ column vector of the $\boldX^{(*)}$, i.e., $2\times1$ column vector containing the physical positions of the $i^{\text{th}}$ magnetic probe in $R$ and $Z$ coordinate. $\sigma_n^2$ is a small number for the numerical stability during matrix inversion,\cite{Kwak:2017gy} and $\delta_{ij}$ is the Kronecker delta. Hyperparameters $\sigma_f^2$, $\ell_R$ and $\ell_Z$ are the signal variance and the length scales in $R$ and $Z$ directions, respectively. These hyperparameters govern the characteristic of the Gaussian process, i.e., Eq. (\ref{eq:gp}), and we select the hyperparameters such that the evidence $p(\boldB)$ is maximized\cite{Kwak:2016kv} with an assumption\cite{Li:2013apa} of $\ell_R=\ell_Z$ for simplicity. As searching for the hyperparameters may become time consuming, thus not applicable for real-time control, one can obtain these values beforehand using many existing plasma discharges as for the case of density reconstruction.\cite{Kwak:2017gy} Once we have values for the hyperparameters, we use Eq. (\ref{eq:gp}) to obtain the values of the missing magnetic signals $\boldB^*$, i.e., $\boldB^*=\matK^*\matK^{-1}\boldB$.
 
\begin{figure}[t]
    \centering
    \includegraphics[width=.85\linewidth]{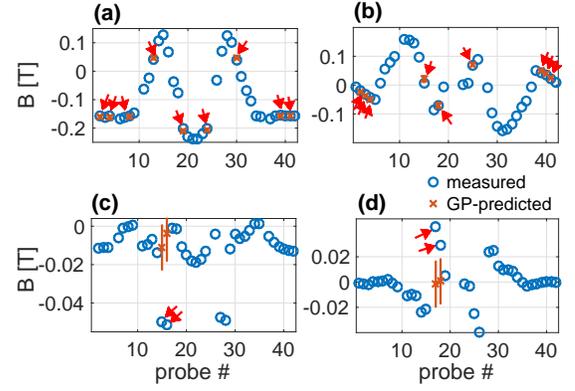}
    \caption{Successful GP predictions (red crosses) compared with the actual data (blue circles) for (a) $B_t$ and (b) $B_n$ at $3.70$\:sec of KSTAR shot \#9010 where we remove nine signals (indicated by red arrows) simultaneously to examine the proposed GP imputation scheme. On the other hand, if the magnetic signals are spatially varying fast such as (c) $B_t$ from MPs \#15 and \#16 and (d) $B_n$ from MPs \#17 and \#18, the GP imputation scheme fails to infer the correct values.}
    \label{fig:MS-gp}
\end{figure}

Fig. \ref{fig:MS-gp}(a) and (b) show that our proposed GP imputation scheme successfully infers the missing magnetic signals both for (a) $B_t$ and (b) $B_n$ where the red crosses are the inferred values and the blue circles are the measured (actual) values. We have examined our scheme with up to nine missing signals indicated by the red arrows. 

We have also found that the GP imputation scheme fails to infer the correct values if the magnetic signals are varying fast in space as shown in Fig. \ref{fig:MS-gp}(c) for $B_t$ and (d) for $B_n$. This is the limitation of the GP-only imputation scheme.

\subsection{Based on Bayes' model coupled with Gaussian Process}
\label{subsec:bayes_GP}

As we recognize the limitations of the Bayes' model (infinite number of solutions for more than one missing magnetic signals of the same component) and the GP (incorrect inference for spatially fast-varying missing magnetic signals), we resolve such weaknesses by combining the two schemes while retaining their advantages. We let the GP finds the values of the missing magnetic signals while satisfying the Maxwell's equations.

Let us, first, select one missing magnetic signal among the missing ones denoted as $B_k^*$, and define $\check\boldX^*$ to contain the positions of $R$ and $Z$ for all the missing magnetic signals except the ones corresponding to $B_k^*$ resulting in $2\times (N_m -1)$ matrix; while $\check\boldX$ containing those of $B_k^*$ in addition to intact magnetic signals becoming $2\times (N_i +1)$ matrix, i.e., concatenate those of $B_k^*$ at the last column of $\boldX$. With $\check\boldX^*$ and $\check\boldX$ our covariance matrices become
\begin{eqnarray}
\label{eq:gpcond_bayes}
&& \quad \check\matK \equiv \check\matK\left(\check\boldX, \check\boldX \right), \quad ( (N_i+1)\times (N_i+1)\text{ matrix} ) \nonumber \\
&&\quad \check\matK^* \equiv \check\matK\left(\check\boldX^*, \check\boldX \right), \quad ( (N_m-1)\times (N_i+1)\text{ matrix} ) \nonumber \\
 &&\quad\check\matK^{**} \equiv \check\matK\left(\check\boldX^*, \check\boldX^* \right), \quad ( (N_m-1)\times (N_m-1)\text{ matrix} ). \nonumber
\end{eqnarray}
From $(N_m-1)\times (N_i+1)$ matrix of $\check\matK^*\check\matK^{-1}$ we separate out the last column and denote this column vector as $\boldL$ and the rest of the matrix, i.e., without the last column of $\check\matK^*\check\matK^{-1}$, be $\matLambda$. Since we have found that $\boldB^*=\matK^*\matK^{-1}\boldB$ in Sec. \ref{subsec:GP}, we obtain
\begin{equation}
\label{eq:gp_reduction}
\check\boldB^*_\oplus = \matLambda\boldB_\oplus + \boldL B_{k\oplus}^*,
\end{equation}
stating that once one missing magnetic signal $B_k^*$ is determined, then all the other missing magnetic signals $\check\boldB^*$ are determined by the GP. We find the unknown $B_k^*$ using the Bayes' model where it is perfectly applicable if we have only one missing signal as discussed in Sec. \ref{subsec:bayes}. Thus, $\boldlambda^{*T}\boldB_\oplus^*$ in Eq. (\ref{eq:likebi}) is
\begin{eqnarray}
\boldlambda^{*T}\boldB_\oplus^* &=& \lambda_k^*B_{k\oplus}^* + \check\boldlambda^{*T}\check\boldB_\oplus^* \nonumber \\ 
&=&\left(\lambda_k^* + \check\boldlambda^{*T}\boldL \right) B_{k\oplus}^* + \check\boldlambda^{*T}\matLambda\boldB_\oplus,
\end{eqnarray}
where $\lambda_k^*$ and $\check\boldlambda^*$ are the segment distances for the selected missing magnetic signal $B_{k\oplus}^*$ and for the rest of the missing signals, respectively.

Slightly modifying Eq. (\ref{eq:likebi}) to include the GP scheme, our likelihood for the Bayes' model, then, becomes
\begin{eqnarray}
\label{eq:likebi_gp}
p( \boldB_\oplus | B_{k\oplus}^*, \Omega_\oplus) &=& \frac{1}{\sqrt{2 \pi} \sigma} \exp \left[ \frac{1}{2\sigma^2} \left(  \left(\lambda_k^* + \check\boldlambda^{*T}\boldL \right) B_{k\oplus}^*  \right.\right.  \nonumber \\
& - & \left.\left. \left( \Omega_\oplus - \boldlambda^T\boldB_\oplus -  \check\boldlambda^{*T}\matLambda\boldB_\oplus  \right) \right) ^2 \right].
\end{eqnarray}
The likelihood now contains only one unknown $B_k^*$, and all the rest of the missing signals are treated as known ones using the GP, i.e,. Eq. (\ref{eq:gp_reduction}). 

We contruct the prior $p(B_{k\oplus}^*|\Omega_\oplus)$ to follow a Gaussian distribution with the mean of $B_{k\oplus}^\text{pair}$ and the variance of $\sigma^2_{\text{prior}}$. $B_{k\oplus}^\text{pair}$ is the signal of the magnetic probe from the up-down symmetric position of the missing signal $B_{k\oplus}^*$, i.e., with the same $R$ and the opposite $Z$. MPs \#6 and \#37, MPs \#12 and \#31, and MPs \#19 and \#24 in Fig. \ref{fig:MS-modelling}(a) are pairs of the up-down symmetric magnetic probes as examples. We use such a paired magnetic signal as a prior mean of the missing signal because KSTAR discharges are quite up-down symmetric, so that a typical correlation between the paired signals is about $0.9$. Regarding the prior variance $\sigma_\text{prior}^2$, to minimize possible biases we set it to be $500$ which means that the prior distribution is largely uniform since the actual values of the magnetic signals are not much larger than $0.1$\:T as shown in Fig. \ref{fig:MS-gp}.  

We finally obtain the posterior following Eq. (\ref{eq:msbayeinf}) as
\begin{eqnarray}
\label{eq:posterior_gp}
p(B_{k\oplus}^* | \boldB_\oplus, \Omega_\oplus) &\propto&   \nonumber \\
&&\!\!\!\!\!\!\!\!\!\!\!\!\!\!\!\!\!\!\!\!\!\!\!\!\!\!\!\!\!\!\!\!\!\!\!\!\!\!\!\!\!\! \exp \left[ - \frac{\left( B_{k\oplus}^* - B_{k\oplus}^\bigstar \right)^2 }{2\sigma_\text{GP}^2} - \frac{\left(B_{k\oplus}^* - B_{k\oplus}^\text{pair}  \right)^2}{2\sigma^2_{\text{prior}}}  \right],
\end{eqnarray}
where 
\begin{eqnarray}
B_{k\oplus}^\bigstar &=&\frac{ \Omega_\oplus - \boldlambda^T\boldB_\oplus -  \check\boldlambda^{*T}\matLambda\boldB_\oplus }{\lambda_k^* + \check\boldlambda^{*T}\boldL }   \nonumber \\
\sigma_\text{GP}^2&=&\left[\frac{\sigma} {\lambda_k^* + \check\boldlambda^{*T}\boldL }\right]^2. \nonumber
\end{eqnarray}
Thus, \textit{maximum a posterior} (MAP) denoted as $B_{k\oplus}^\text{MAP}$ can be analytically estimated and is
\begin{equation}
\label{eq:bkmap}
B_{k\oplus}^\text{MAP}=\left[\frac{B_{k\oplus}^\bigstar}{\sigma_\text{GP}^2} + \frac{B_{k\oplus}^\text{pair}}{\sigma_\text{prior}^2} \right]\left[\frac{1}{\sigma_\text{GP}^2} + \frac{1}{\sigma_\text{prior}^2} \right]^{-1}
\end{equation}
with the posterior variance $\sigma^2_\text{post} = (1/\sigma_\text{GP}^2 + 1/\sigma_\text{prior}^2 )^{-1}$. Once $B_{k\oplus}^\text{MAP}$ is found, then all the other missing signal are determined by Eq. (\ref{eq:gp_reduction}). This completes the imputation process.

\begin{figure}[t]
    \centering
    \includegraphics[width=0.485\textwidth]{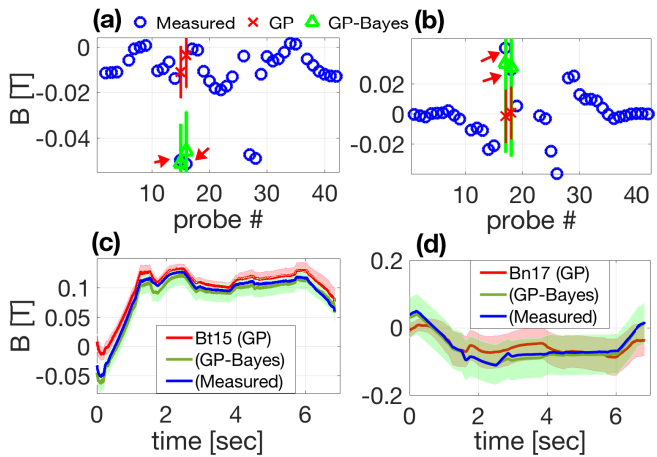}
    \caption{(a) $B_t$ from MPs \#15 and \#16 and (b) $B_n$ from MPs \#17 and \#18 from KSTAR shot \#9010 at $0.1$\:sec as shown in Fig. \ref{fig:MS-gp}(c) and (d). Green triangles obtained by the Bayes' mode with the GP matches the measured values (blue circles) well, while the GP-only method (red crosses) fails to do so as has been discussed in Sec. \ref{subsec:GP}.  Comparisons of temporal evolutions for (c) $B_t$ from MP \#15 and (d) $B_n$ from MP \#17 where blue line is the measured values, red line for the GP-only and green line for the Baye's model with the GP. Green lines agree well with blue lines well throughout the whole discharge including ramp-up and ramp-down phases.} 
    \label{fig:MS-gpwithbayes}
\end{figure}

To validate our proposed imputation scheme based on the Bayes' model with the Maxwell's equations coupled with the GP, we take the same examples shown in Fig. \ref{fig:MS-gp}(c) and (d). Fig. \ref{fig:MS-gpwithbayes}(a) $B_t$ from MPs \#15 and \#16 and (b) $B_n$ from MPs \#17 and \#18 from KSTAR shot \#9010 at $0.1$\:sec show considerable improvements where the green triangles inferred by the Bayes' model coupled with the GP are very close to the blue circles which are the measured values. Again, the red crosses obtained only by the GP fails to do so. 

Fig \ref{fig:MS-gpwithbayes}(c) $B_t$ from MP \#15 and (d) $B_n$ from MP \#17 from KSTAR shot \#9427 show temporal evolutions of the inferred values where the blue line is the measured values, the red line for the GP only and the green line for the Bayes' model with the GP. Typically, the GP-only method fails largely during ramp-up and ramp-down phases while it is not too bad during the flat-top phase; whereas the Bayes' model with the GP finds the correct values throughout the whole discharge.

Eq. (\ref{eq:bkmap}) contains no unknowns which means that $B_{k\oplus}^\text{MAP}$ can be estimated in real-time. In fact, our proposed method takes less than $1$\:msec on a typical personal computer. The hyperparameters are prepared beforehand based on many previous discharges, and missing or faulty signals can be identified\cite{Neto:2014cm, Nouailletas:2012ho} in real-time. What one requires to do is simply to perform the following three steps in real-time: (1) select a missing signal ($B_{k\oplus}^*$) among all the missing ones ($\boldB_\oplus^*$), (2) estimate noise levels ($\sigma$) of the measured signals and (3) apply Eq. (\ref{eq:bkmap}) and Eq. (\ref{eq:gp_reduction}) to impute more than one missing magnetic signals. Good choice of a missing signal ($B_{k\oplus}^*$) is from the ones that spatially vary fast if they exist. In KSTAR such signals are $B_t$ from MPs \#15 and \#16, and $B_n$ from MP \#17 and \#18 in almost all cases, if not all.

\section{Conclusion}
\label{sec:conclusion}

We have developed and presented a real-time inference scheme, thus imputation scheme, for missing or faulty magnetic signals. Our method, Bayes' model with the likelihood constructed based on the Maxwell's equations, specifically Gauss's law of magnetism and Amp\`ere's law, coupled with the Gaussian process, allows one to infer the correct values even if more than one missing signals that are spatially varying fast exist, outperforming the Baye's-only and the GP-only methods without losing their own advantages. We have examined our method up to nine missing magnetic signals.

The proposed method takes less than $1$\:msec on a typical personal computer, so that the method can be applied to fusion-grade plasma operations where real-time reconstruction of magnetic equilibria is crucial. It can also be used for a neural network trained with a complete set of magnetic signals without fearing the possible loss of magnetic signals during plasma operations.

As a possible future work, developing a real-time searching algorithm for the hyperparameters in the Gaussian process that optimizes the evidence will be beneficial. Although the current results with the predetermined hyperparameters based on many previous discharges are satisfying, the hyperparameters specific to a current discharge may provide much better plasma controls especially for those discharges that we have not yet explored much.

\section*{Acknowledgement}
This research was supported by National R\&D Program through the National Research Foundation of Korea (NRF) funded by the Ministry of Science and ICT (grant number NRF-2017M1A7A1A01015892 and NFR-2017R1C1B2006248) and the KUSTAR-KAIST Institute, KAIST, Korea.

\section*{References}

\bibliographystyle{apsrev4-1}
\bibliography{jpbib_imp}

\end{document}